\documentclass[12pt]{article}
\usepackage{amsmath,amssymb,amsthm}
\usepackage{latexsym,graphicx,color,subfigure,mathrsfs}
\usepackage{natbib}
\usepackage{epstopdf}

\def\Tr{{\rm Tr}}

\def\p{\partial}

\def\D{\mathcal{D}}

\def\hsp{\, ,\hspace{.7cm}}
\def\=:{=\hspace{-.7em}\raisebox{1.1ex}{.}\hspace{.1em}\raisebox{-0.2ex}{.} }

\newcommand {\beq}{\begin{eqnarray}}
\newcommand {\eeq}{\end{eqnarray}}

\def\diag{{\rm diag}}

\theoremstyle{definition}

\makeatletter
\@addtoreset{equation}{section}

\renewcommand{\thefootnote}{\fnsymbol{footnote}}
\makeatother

\makeatletter
\newcommand{\thetablename}{Table}
\def\fnum@table{\thetablename\ \thetable}
\makeatother

\begin{document}

\thispagestyle{empty}

\begin{center}
{\Large \bf Dwarf Galaxy Sized Monopoles as Dark Matter?}  
\\[15mm]
{Jarah~{\sc Evslin}}$^{1}$\footnote{jarah(at)ihep.ac.cn} and 
{Sven Bjarke~{\sc Gudnason}}$^{2}$\footnote{gudnason(at)phys.huji.ac.il}
\vskip 6 mm

\bigskip
{\it
$^1$ TPCSF, Institute of High Energy Physics, CAS, P.O.~Box 918-4,
Beijing 100049, P.R.~China\\
$^2$ Racah Institute of Physics, The Hebrew University,
Jerusalem 91904, Israel\\
}

\bigskip

\bigskip

\bigskip

\bigskip
\bigskip
\bigskip

{\bf Abstract}\\[5mm]
{\parbox{15cm}{\hspace{4mm}
\small

\noindent
We propose a model of dark matter: galaxy-sized 't Hooft-Polyakov
magnetic monopoles in a new, extraordinarily weakly coupled SU(2)
gauge sector with an adjoint Higgs field and two flavors of
fundamental fermions.    We fit the parameters by asserting that the
dark matter halos of the lightest dwarf spheroidal (dSph) galaxies
consist of a single charge $Q=1$ monopole.  Lensing and wide binary bounds are then
easily satisfied and the monopoles form in time to help with CMB
fluctuations. In this model dSph and low surface brightness (LSB)
halos automatically have (1) 
A minimum mass - Dirac quantization solves the missing satellite
problem, (2) A constant density core $(r<r_1)$, (3) An intermediate
regime $(r_1<r<r_2)$ with density $\rho\sim 1/r^2$.  The model
predicts that (A) $r_1$ is proportional to the stellar
rotational/dispersion velocities at $r_1<r<r_2$, (B) $r_2$ is
reasonably $Q$ independent and so dSph halos extend at least ten times
farther than their half-light and tidal radii, (C) The minimal stellar
dispersion is $1/\sqrt{2}$ times the next-smallest allowed value.  A serious potential problem with our proposal is that non-BPS monopoles are repulsive. The Jackiw-Rebbi mechanism yields four species of monopoles, and we assume that, for some choice of Yukawa couplings,
one species is light and serves only to screen the repulsive
interactions of another.

}}
\end{center}
\newpage
\pagenumbering{arabic}
\setcounter{page}{1}
\setcounter{footnote}{0}
\renewcommand{\thefootnote}{\arabic{footnote}}

\section{Introduction}

\subsection{Motivation}

Dwarf spheroidal (dSph) and low surface brightness (LSB) galaxies provide the
purest known samples of dark matter in the Universe, and so provide a
unique laboratory for studying its properties.  The CMB indicates that
the Universe was much more homogeneous before recombination than it is
today, thus dark matter must be made of something 
which is able to clump together in a reasonably short amount of time 
(13 billion years).  This has led to a wide acceptance that dark
matter consists of particles which are so massive that they were
already moving at nonrelativistic speeds during recombination, such
particles are referred to as cold dark matter (CDM).

Simulations of structure formation in the presence of cold dark matter
generally agree on two predictions. 
First, there should be hundreds if not tens of thousands of dwarf
satellite galaxies in our local group \citep{klypin,moore1} with masses
of under $10^6\, M_\odot$.  Second, the density profiles of these
galaxies should be qualitatively of the form proposed by Navarro,
Frenk and White \citep{NFW}, in particular they should have cusps at
small radii $r$, diverging at least as $1/r$, with more recent studies
suggesting $1/r^{3/2}$ \citep{moore2}.  As we now review, neither of
these predictions agrees with observations.

Recently the number of known dwarf galaxies has been increasing quite
rapidly, but various studies still place it between 4 and 400 times
below the cold dark matter prediction in our local group \citep{mateo}
and beyond \citep{2011nani}. On the contrary,
galactic masses appear to have a lower bound.  More specifically, the
lightest satellite galaxies of the Milky Way all appear to have about
$10^7\, M_\odot$ within 300 pc\footnote{ Only two dwarf
  galaxies (Coma Berenices \citep{comaber} and Segue I
  \citep{segueuno}) have been observed with masses compatible with
  $10^6\, M_\odot$ or less.  However, due to their close proximities
  to the Milky way, these dSphs have half-light radii of about 60 and
  30 pc respectively \citep{Belokurov}.  The measured values
  of the mass reflect only the mass inside of this radius, and so are
  very consistent with $10^7\, M_\odot$ within 300 pc.} of
their center \citep{Kap1000} and between $10^7$ and $10^8\, M_\odot$
within 600 pc \citep{Strigari2000}.  Thus the light dark
matter halos predicted by CDM simulations have not been observed.

This is not in itself a contradiction.  It is a logical possibility
that halos exist with all of the masses predicted by CDM, but those
lighter than this minimum mass do not have deep enough gravitational
potentials to have led to star formation, as a consequence of gas
being lost to radiation during reionization
\citep{reionization,reionization2}, 
supernova feedback \citep{snfeedback} or cosmic ray pressure
\citep{wadepuhl}, as has been supported for example by the Aquarius
simulations in \citet{aquariustuttobene}.  The problem
\citep{problemaNFW,problema} is that the 6 halos simulated by the
Aquarius project \citep{aquarius} as well as the halo simulated by Via
Lactea II \citep{latte} also each have at least 10 halos in the mass
range between that of dSphs and irregular dwarfs, whereas no dwarfs in
this mass range have been found in our local group.  While it is
plausible that the missing light satellites have not been seen because
they simply lack stars, no explanation has been proposed for the
missing heavy satellites.

In \citet{NFW} it was already shown that density profiles of dwarf 
galaxies do not exhibit divergent 
cusps, on the contrary they are consistent with constant density cores.
In the case of dSphs, a core with a reasonably constant density is
required both by the long lifetimes of globular clusters embedded in
dwarfs \citep{kleyna,goerdt,0703308} and also by analyses of
chemically distinct stellar components \citep{1108.2404}. Such a
constant density core appears in the King model \citep{king} of a
steady state system, however dSph galaxies have not had time to settle
into such a steady state \citep{0703308} and also the model poorly
captures the observed stellar kinematics \citep{wu}.   

Beyond the core,
dark matter dominated galaxies exhibit an intermediate
radius regime $r_1<r<r_2$ with density $\rho\propto 1/r^2$.  This can
be seen clearly in the HI (the 21 cm 1S hyperfine transition) rotation
curves for example for the 15 fairly large dwarf galaxies and LSBs
presented in \citet{swaters}.  The corresponding H$\alpha$ ($n=3$ to
$n=2$ transition) curves probe the central regions of these galaxies
and are consistent with constant density cores.  In
\citet{walker}, stellar velocity dispersions in the intermediate
regimes of 8 dSphs are plotted.  These dispersions are
radius-independent, and so the Jeans equation again supports a $1/r^2$
density profile.  If the intermediate region extended to infinite
radius, the halo mass would be linearly divergent and so the finite
mass implies the existence of a third regime $r>r_2$ in which the
density falls faster than $1/r^3$.

\subsection{Monopoles}

Summarizing, dark matter dominated galaxies, in contradiction with CDM
simulations, have a minimum mass, a constant density core, an
intermediate regime with $1/r^2$ density and an external regime in
which the density falls faster.   While particulate models of dark
matter may fail to reproduce these facts, in this note we claim that
they are reproduced by a new model of dark matter in which each
galactic halo consists of a single, extended particle called a 't
Hooft-Polyakov monopole \citep{thooft,polyakov}.  These particles are
classical field theory solutions.  They are stable, and as a result of
Dirac quantization they are automatically quantized, meaning that
there is a lightest solution.  We will identify the lightest solution
with the dark matter halo of the lightest dSph.  Therefore in our
model, the missing (light) satellite problem is automatically solved by the fact
that any lighter halo would violate the Dirac quantization condition,
and so not correspond to any finite energy solution of our quantum
field theory.   We will review the fact that the density profiles of
these solutions have just the same structure as those observed in dark
matter halos. 

Our model will not require any radically new physics.  We just add to 
the standard model a new sector which is roughly the model of
electroweak interactions proposed in
\citet{gg}.  More precisely, we consider a new SU(2) gauge theory
with an adjoint scalar field and two fundamental spinor
fields\footnote{We would like to emphasize that this SU(2) gauge group
  has nothing to do with the SU$(2)_L$ of the electroweak theory of
  the standard model. It is a new gauge group and the new matter
  fields are neutral with respect to the standard model gauge 
  symmetries. This dark sector interacts with the standard model
  gravitationally, and we leave open the possibility that standard
  model fields may be charged under the dark gauge symmetry. }.  We
consider a Higgs potential for the scalar field with quartic coupling
larger than the square of the gauge coupling and so the resulting
monopoles will be highly non-BPS.   
  
The bosonic sector of the model, which is all that will be relevant
for the smallest dSphs and for galactic cores, has only 3 free
parameters: the gauge coupling, the tachyonic mass of the Higgs field
and its quartic coupling.  These three parameters need to reproduce
the rotation curves of all dark matter dominated galaxies, and so the
model is overconstrained and very falsifiable.  Also, given these
three parameters, one can determine when the monopoles formed via the
Higgs mechanism and also whether the gauge interactions dominate over
gravitational interactions inside of the halos.  Consistency with CMB
perturbations requires\footnote{We thank Malcolm Fairbairn for
  stressing this point.} that the monopoles nucleated before the last
scattering surface, and the relevance of the gauge theory solutions
requires gauge interactions to dominate over gravity inside of the
halo.  We will see that the three parameters determined using galactic
rotation curves fall within the reasonably narrow window that
satisfies both of these conditions, and thus the model surprisingly
passes two nontrivial consistency tests. 

Our monopoles are far heavier than the upper bounds on MACHO masses
excluded by gravitational lensing searches \citep{macho}.   Traditional
MACHO models introduce inhomogeneities at distance scales of parsecs
or less and so are ruled out by studies of wide binaries
\citep{binari}, but our monopoles are so spatially extended and
homogeneous at these scales that they survive these tests as
well.

\subsection{Screening}

Despite this list of nice properties, there is one major problem with
a model of galactic dark matter halos using non-BPS 't Hooft-Polyakov
monopoles of various charges $Q$.  The problem is that for charges $Q$
greater than $1$, the monopoles are unstable, decaying into monopoles 
of charge $1$.  The reason is that the scalar field mediates an
attractive force, but for all positively charged monopoles the U(1)
magnetic fields repel.  As the scalar is massive and the photon is
massless, the magnetic fields win at large distances and the galaxies
repel, and even explode.  This is certainly inconsistent with
observations. 

This is similar to a world made only of protons\footnote{A crucial
  difference is that in the case of our monopoles, the repulsion
  results solely from the field configuration far from the monopole,
  therefore a screening mechanism in this outer region may suppress
  the unwanted repulsion while leaving the inner and intermediate
  layers intact and well described by the bosonic sector of the field
  theory.}.  Perhaps protons near enough to
each other could form nuclei, due to the short distance attractions of
the strong force (in this case the scalar field and gravity), but at
large distances there would be a larger and larger electric field
leading to repulsion and even Olber's paradox.  The solution to this
problem is that the Universe has as many electrons as protons.  The
electrons screen the protons, as they have opposite charges.  Yet the
protons and electrons do not annihilate each other, as they possess
distinct conserved charges.  If the electrons are free, these
configurations are plasmas or jellium.  If the electrons are bound,
the configurations are atoms.  If the electrons are light enough, they
may condense and the screening will resemble that of a superconductor.
In any case, the magnetic repulsion between distant protons is
screened, and perhaps the repulsion between nearby protons is tempered
enough to allow the attractive forces to win. 

How can this screening be realized in the case of magnetic monopoles?
We need magnetic monopoles with different conserved charges, and
opposite magnetic charges.  We will consider the case in which the
differing charges correspond to global symmetries, to minimize the new
interactions that need to be introduced.  There is only one known
mechanism that allows magnetic monopoles to acquire global charges,
the Jackiw-Rebbi mechanism \citep{JR}.  For each flavor of fundamental
fermions in the theory, the magnetic monopoles acquire a fermion
number of $\pm 1/2$ with respect to that flavor.  Thus in the case of
$2$ flavors there will be $2$ binary choices of charges, and so $4$
possible charges corresponding to $4$ kinds of monopoles and their
antimonopoles.  We will consider monopoles of one flavor and an equal 
number of antimonopoles of another, so that we have an equal number of
positive and negative charges. 

The fermion wave functions are stabilized by the Yukawa couplings.  In 
\citet{JR}, the authors considered Yukawa couplings of order the
gauge coupling $g$.  In this case the fermionic wave functions are
subdominant by a factor of $g$, which we will see is extraordinary
small in our models and so the fermionic contribution is irrelevant.
The monopoles will therefore all have similar masses, and our
monopoles and antimonopoles may annihilate leaving behind the much
lighter fermions.  Different species of galaxies which annihilate into
particles have not been observed, and so this cannot be the right
approach.  Instead we will consider Yukawa couplings of
$\mathcal{O}(1)$, leading to fermionic wave functions of the same order
as the bosonic wave functions.  This opens the possibility that one of
the fermionic flavors is light.  Perhaps, as in the $r$-vacua of
supersymmetric gauge theories \citep{SW2,CKM}, this means that only
monopoles of particular flavors condense, or at least only one becomes
light.  Once one of the flavors is light, like the electron, whether
it condenses or not, it will not attract enough stars to yield a new
flavor of galaxy and so it may provide a reasonable screening
candidate.  Hopefully in this case the interflavor monopole
annihilation cross section is suppressed.  Of course, the fermions
will also yield an $\mathcal{O}(1)$ correction to the heavy monopole
wave function.  Thus there will be systematic errors of
$\mathcal{O}(1)$ in this entire note.  A next logical step would be to
make a concrete choice of fermion couplings, adjusting them to make
one monopole light, and to attempt to understand the resulting
corrections on the second monopole.  There will be further corrections
to the profile, resulting from the screening itself, in the outer
regions of the massive monopoles $r\gtrsim r_2$.  Fortunately we will
be able to learn a lot about our model from smaller radii where the
effect of screening can be safely neglected, which is also the region
in which the dark matter profile is most constrained by observations.

We begin in Sec.~\ref{predsez} with a summary of the key predictions
of this model.  In Sec.~\ref{teoria} we introduce the gauge theory and
the monopole Ansatz.  We will ignore the fermions entirely,
considering only the massive monopole flavor.  We will review a
combination of analytic and numerical results on non-BPS charge $Q=1$
monopoles, in particular displaying the density profiles in the $3$ regions described
above. In Sec.~\ref{valori} we provide formulae for the parameters of
the gauge theory in terms of observables of these lightest galaxies
and estimate their numerical values.   Next in Sec.~\ref{Q} we provide
a rough scaling for the $Q$-dependence of the monopole solutions,
ignoring the intermonopole repulsion and screening.  We check to see
whether the parameters of the theory, already fixed in
Sec.~\ref{valori}, are able to reproduce the rotation curves of higher
$Q$ galaxies.  In Sec.~\ref{gravita}, the Newtonian gravitational
potential is calculated and the single ($Q=1$) monopole theory is also
coupled to general relativity.  In both cases gravitational effects
are seen to be negligible.  Indeed in this model the $1/r^2$ density
profile results not from gravity, but from the topology of the Higgs
field. 

\section{Predictions} \label{predsez}

Needless to say, our model suggests that direct and indirect WIMP searches will not find cosmologically significant quantities of dark matter.  It also yields at least three astrophysical predictions.  First, the core radius is
proportional to the flat velocity dispersion and rotational velocities
in the intermediate range, which we will see are both proportional to
the square root of the charge $r_1\propto vel\propto\sqrt{Q}$.
Second, just as Dirac quantization solves the missing satellite problem by
implying that no halos exist with masses between those of globular
clusters $(Q=0)$ and the smallest dSphs $(Q=1)$,  it also implies that
the minimal velocity dispersion for a dSph is smaller than the second
minimal value by a factor of $\sqrt{2}$.  For example the 6 km/s
average dispersion \citep{walker} in Sextans and Carina may correspond
to $Q=1$, and the 9 km/s of the Ursa Minor dwarf to $Q=2$.  An
improvement in these measurements by a factor of 3 or 4 may already be
enough to confirm or refute this prediction and therefore the model.   

Finally, we will see that a consistent screening mechanism, in which
galaxies only interact with each other gravitationally, requires $r_2$
to be reasonably independent of $Q$.  This implies that dark matter
halos of dSphs extend more than an order of magnitude beyond both
their tidal radii and their half-light radii.  Extending beyond their
tidal radii is not problematic, as the halos are held together not by
gravity but by the interactions in the new gauge sector.    How would
one detect such an extension of the halo radius?  First of all, one
would expect stars beyond the tidal radius to be stripped away more
slowly, leaving more matter beyond the tidal radius than would
otherwise be expected, agreeing with the observations of
\citet{tidal}.  There have been numerous studies of the stars
beyond the tidal radius of the Fornax dwarf, with some suggesting that
they result from simple accretion \citep{battifornax} and some
suggesting that it provides evidence for such a large dark matter halo
\citep{fornaxhalo}.  It would also affect the structure of objects
created by tidal forces such as the Sagittarius dwarf's large
tail \citep{sagittarius} or the change in slope of stellar density
distributions at the tidal radius measured in
\citet{tidalprofiles}.   Effects of the halo size on the evolution
of the stellar dispersions can be used to estimate the mass of the
halo during the period in which it formed.  Such an estimate, using
the Jeans equation at a radius at which the effects of anisotropy are
minimal, was provided in \citet{walker,wolf}.  They found a
favored value of order $3\times 10^9$\ $M_\odot$ for the original
masses of dSph halos, which fits the current halo mass that would be
predicted in our model to within a factor of two. 

If the intermediate regions of dSphs are indeed an order of magnitude
larger than expected, then one would expect them to contribute an
order of magnitude more to gravitational lensing than expected, which
may explain the results of \citet{lensing1,lensing2} which appear
to find much more lensing caused by dwarfs than can be accounted for
from using their known abundances and masses within their half-light
radii.   

While all dwarf galaxies that have been observed appear to have a
minimum mass, our model predicts much more.  All dark matter halos,
even those not containing stars, must also contain this minimum mass,
in sharp contradiction with CDM predictions.  Furthermore, while large
scale structure is known to form hierarchically from smaller
components, this minimum halo mass must be respected during all times
since the formations of the halos themselves, in other words this
model would be falsified by a single lower mass halo even at high
redshift.  Gravitational lensing may be used to search both for halos
which contain no stars and also for halos at high redshift, and thus
provides a powerful tool for falsifying this model.  In particular,
the technique of \citet{lenstec} can be used to detect dwarf
galaxies an order of magnitude lighter than the lightest observed, and
at least two dwarfs have so far been observed using this technique,
one at redshift $z=.22$ \citep{z22} and another at redshift $z=.81$
\citep{z81}.   The measurement of these masses depends strongly on the
profile template used, and needless to say the profile used is quite
different from the halo profile suggested here, however in both cases
the mass measured within 600 pc of the core is above the
minimum dSph mass observed in our local group, and so is consistent
with the absolute minimum mass required by our model.  A similar
technique with an identical template was used to find one or two dwarf
galaxies at $z=.46$ in \citet{z46}.  The second dwarf, whose
existence is quite uncertain, has a mass which is compatible with that
of the lightest dwarfs but is also compatible with a lower mass.
Therefore a more precise measurement of this galaxy can potentially
lead to a time-dependent minimum mass, which would falsify our model.
If this technique could be applied to determine the size of the halo,
and if this size exceeds the tidal radius, it would
provide a smoking gun for our nongravitationally bound halo proposal.

\section{The gauge theory and its monopole solutions} \label{teoria}

We will briefly review the 't Hooft-Polyakov monopole solution in an
SU(2) gauge theory with an adjoint matter field $\Phi$.
We would like to stress again that this gauge group is 
new and is not the same as the SU$(2)_L$ gauge symmetry of the
standard model. 
Furthermore, the fields of the model are neutral with respect to
the standard model charges and so likely only interact with the standard
model particles gravitationally.
The model is described by the Lagrangian density
\beq
\mathcal{L}^M = \Tr\left[-\frac{1}{2g^2}F_{\mu\nu}F^{\mu\nu}
-\D_\mu \Phi \D^\mu \Phi\right]
-\frac{\lambda}{4}\left(v^2 -2\,\Tr\left[\Phi^2\right]\right)^2 \, , \label{gg}
\eeq
where $\Phi=\Phi^a t^a$ is an $\mathfrak{su}(2)$ algebra-valued
adjoint Higgs field, $g$ is the gauge coupling, $\lambda>0$ is the scalar
coupling, $\D_\mu\Phi=\p_\mu\Phi+i[A_\mu,\Phi]$ and finally 
$F_{\mu\nu}=\p_\mu A_\nu - \p_\nu A_\mu + i[A_\mu,A_\nu]$ is the
Yang-Mills field strength tensor. We use the convention 
$\Tr[t^a t^b]=\delta^{ab}/2$.
The equations of motion are
\beq
\D_\mu\D^\mu\Phi = 
-\lambda \left(v^2 - 2\,\Tr\left[\Phi^2\right]\right)\Phi \, , \qquad
\D_\mu F^{\mu\nu} = -i g^2 \left[\D^\nu \Phi, \Phi\right] \, . 
\eeq
The (hedgehog) radial Ansatz for a single (charge $Q=1$) regular 't
Hooft-Polyakov monopole reads 
\beq
\Phi = \frac{1}{r} v h(r)x^a t^a \, , \qquad
A_i = -\frac{1}{r^2}\left(1-k(r)\right)\epsilon_{ija} x^j t^a \, ,
\label{eq:Ansatz} 
\eeq
where $r$ is the (spatial) radial coordinate.  The functions $h(r)$
and $k(r)$ parametrize the Higgs field and gauge field respectively
and must solve the equations of motion
\begin{align}
h'' + \frac{2}{g v r}h' &= \frac{2}{(g v r)^2}k^2 h 
  - \frac{\lambda}{g^2} \left(1-h^2\right)h \, , \label{eq:EOM1} \\
k'' &= \frac{1}{(g v r)^2}\left(k^2 - 1\right)k + h^2 k \, ,
  \label{eq:EOM2}
\end{align}
where ${}'\equiv \frac{d}{d(gvr)}$ is a rescaled radial
derivative. The reason for writing the equations of motion in this
form is that it is easy to identify the independent variables;
i.e.~$1/(g v)$ sets the overall length scale and $\lambda/g^2$
characterizes the non-BPS-ness of the system. $\lambda = 0$ is the BPS
case and $\lambda/g^2 \gg 1$ is strongly non-BPS.

In a static configuration the energy density is
\begin{align}
\mathcal{H} = 
\frac{1}{g^2}\left(\frac{k'}{r}\right)^2 
+\frac{1}{2g^2r^4}\left(1-k^2\right)^2
+\frac{1}{2}\left(v h'\right)^2
+\left(\frac{v k h}{r}\right)^2
+\frac{\lambda v^4}{4}\left(1-h^2\right)^2 \, . 
\label{eq:Ansatz_energydensity}
\end{align}
We are interested in configurations with a finite total energy.
Finiteness at small and large radii implies the boundary conditions 
\beq
h(0) = 0 \, , \quad
k(0) = 1 \, , \quad
h(\infty) = 1 \, , \quad
k(\infty) = 0 \, . 
\label{eq:Ansatz_BC}
\eeq

The magnetic charge can be calculated using Gauss' law for the
unbroken U(1)
\beq
Q = \frac{1}{2\pi v} \int_{\mathbb{R}^3} 
\Tr \left[B_i \D_i \Phi \right] 
= \int_0^\infty dr \; \frac{d}{dr}\left(h\left(1-k^2\right)\right) 
= 1 \, , 
\eeq
with $B_i \equiv \frac{1}{2}\epsilon_{ijk}F_{jk}$. The second equality
follows from the hedgehog Ansatz \eqref{eq:Ansatz}, while the
third equality is found using the boundary conditions
\eqref{eq:Ansatz_BC}. This charge is the topological monopole charge,
which is valued in the second homotopy group of the gauge
orbits of the space of Higgs vacua 
$\pi_2(SU(2)/U(1))\simeq \pi_2(S^2)=\mathbb{Z}$.  

Perhaps the simplest and most studied case is that in which the
coupling $\lambda$ is sent to zero while the symmetry breaking is kept
by fixing the VEV of $\Phi$ equal to $v$. The corresponding solution
is the Bogomol'nyi-Prasad-Sommerfield (BPS) monopole
\citep{Bogomolny:1975de,Prasad:1975kr} which satisfies the BPS
equations 
\beq
\frac{1}{g v} h' = \frac{1}{(g v r)^2}\left(1-k^2\right) \, , \qquad
\frac{1}{g v} k' = -h k \, .
\eeq
These equations can be integrated, yielding the analytic solution
\beq 
h = \coth(g v r) - \frac{1}{g v r} \, , \qquad
k = \frac{g v r}{\sinh(g v r)} \, ,
\eeq
which has a mass
\beq 
M_{\rm BPS} = \frac{4\pi v}{g} \, . 
\eeq

The BPS monopole has no intermediate region in which the density
scales as $1/r^2$, and so does not resemble a dark matter halo.  We
will therefore consider a monopole with $\lambda/g^2 \gg 1$.  
The monopole mass is a function of the parameter $\lambda/g^2$
\beq 
M(\lambda/g^2) \simeq \frac{4\pi v}{g} f(\lambda/g^2) \ , 
\label{eq:nonBPSmass}
\eeq
where $f(\lambda/g^2) \in [1,1.79]$ is a smooth and monotonically
increasing function with the limiting values  $f(0)=1$ and
$f(\infty)=1.79$ as shown in \citet{Kirkman:1981ck}.

\section{The monopole halo for the minimal dSph} \label{valori}

\subsection{The approximate density profile}

We now identify the basic $Q=1$ magnetic monopole described above with
the dark matter halo of the lightest dwarf galaxies.  We may express
the parameters of the bosonic sector \eqref{gg} of our theory in terms
of observable properties of these galaxies.   However we stress that
at radii $r\gtrsim r_2$ the screening mechanism is necessarily
important, and thus calculations of the behavior in this regime and
consequently even of the value of $r_2$ itself are dependent on the
specific model describing this screening.  As screening is not
necessary for the stability of the $Q=1$ monopole, we will ignore it
in this section, simply recalling that the value of $r_2$ so derived
cannot be trusted. 

In order to parametrize the structure of the basic monopole, we
will estimate the energy density profile.  We will consider the
following approximation 
\beq
h(r) = 
\begin{cases}
\frac{r}{r_1} \ , & \quad r \in [0, r_1] \, , \\
1 \ , & \quad r \in (r_1, \infty) \, ,
\end{cases} \qquad 
k(r) = 
\begin{cases}
1 \ , & \quad r \in [0, r_2] \, , \\
0 \ , & \quad r \in (r_2, \infty) \, .
\end{cases}
\label{eq:single_crude}
\eeq
In particular we neglect the exponential tail of $k(r)$ at $r>r_2$,
which we expect to be dominated by the model-dependent screening. 

Substituting this estimate into the energy density
\eqref{eq:Ansatz_energydensity} and integrating over all space we find 
\beq
M = \frac{4\pi v}{g}
\left[\frac{1}{2 g v r_2} + g v \left(r_2 - \frac{1}{2} r_1\right)
+ \frac{2\lambda}{105g^2}(g v r_1)^3 \right] \, . 
\label{eq:approx_mass}
\eeq
Now, if we minimize this expression with respect to $r_{1,2}$ we
obtain the following characteristic radii for the single monopole
\beq
r_1 = \frac{1}{2 v}\sqrt{\frac{35}{\lambda}} \hsp
r_2 = \frac{1}{\sqrt{2} g v} \, . \label{raggi}
\eeq
Intuitively these two formulae express the fact that the scalar and
$W$ boson masses are of order $v\sqrt{\lambda}$ and $gv$ respectively, 
and $r_1$ and $r_2$ are their corresponding de Broglie wavelengths. 
When we estimate these quantities, we will see that these masses are 
extremely small, roughly $10^{-25 }$ eV and $10^{-28}$ eV,
respectively.  In a subsequent publication we intend to study the
compatibility of such light particles with cosmological bounds. 

The ratio of the radii is
\beq 
\frac{r_2}{r_1} = \sqrt{\frac{2\lambda}{35g^2}} \, . \label{rr} 
\eeq
This implies that the relative size of the intermediate region with
respect to the core is proportional to $\sqrt{\lambda/g^2}$. The fact
that galactic rotation curves exhibit a large flat region then implies
that $\lambda/g^2$ is large and so our monopoles are very non-BPS.

We can now find the scaling behavior of the density profile in the
various regions using Eq.~\eqref{eq:single_crude} and the energy
density \eqref{eq:Ansatz_energydensity} 
\beq
\mathcal{H}(r) = 
\begin{cases}
\frac{3v^2}{2r_1^2} 
+\frac{\lambda v^4}{4}\Big[1-\big(\frac{r}{r_1}\big)^2\Big]^2
 \ , & \quad r \in [0, r_1] \ , \\
\frac{v^2}{r^2} \ , & \quad r \in (r_1, r_2] \ , \\
\frac{1}{2 g^2 r^4} \ , & \quad r \in (r_2, \infty) \, . 
\end{cases}
\label{eq:energydensities}
\eeq
Note that this approximate energy density is not continuous, which the 
real solution of course is. 
It is however quite clear from this small calculation that there are 
three regimes inside of the monopole: the core which has a quite high
and roughly constant energy density, an intermediate regime where the
energy density drops as $1/r^2$ and finally a tail where the energy
density drops as $1/r^4$, although this last estimate is very
sensitive to the screening.   

Substituting the radii $r_{1,2}$ \eqref{raggi} into the mass
\eqref{eq:approx_mass}, we find
\beq
M = \frac{4\pi v}{g}  
  \left[\sqrt{2} - \frac{1}{6}\sqrt{\frac{35}{\xi}}\right] \ , 
\eeq
which corresponds to $f(\infty) \sim \sqrt{2} = 1.44$ - not that
far from the precise numerical result $f(\infty)\sim 1.79$. 
We can also estimate the mass in each region of the monopole. 
In the core $r<r_1$ 
\beq
M_{r<r_1} = \frac{4\pi v}{g} \frac{1}{3}\sqrt{\frac{35}{\xi}}
\, ,
\eeq
while in the intermediate region $r_1<r<r_2$ it is
\beq
M_{r_1<r<r_2} = \frac{4\pi v}{g} 
\left[\frac{1}{\sqrt{2}} - \frac{1}{2}\sqrt{\frac{35}{\xi}}\right] \, , 
\eeq
and finally for $r>r_2$, where we hope that the screening will 
dramatically alter the profile
\beq
M_{r>r_2} = \frac{4\pi v}{g} \frac{1}{\sqrt{2}} \, .
\eeq

\begin{figure}[!tp]
\begin{center}
\includegraphics[width=0.6\linewidth]{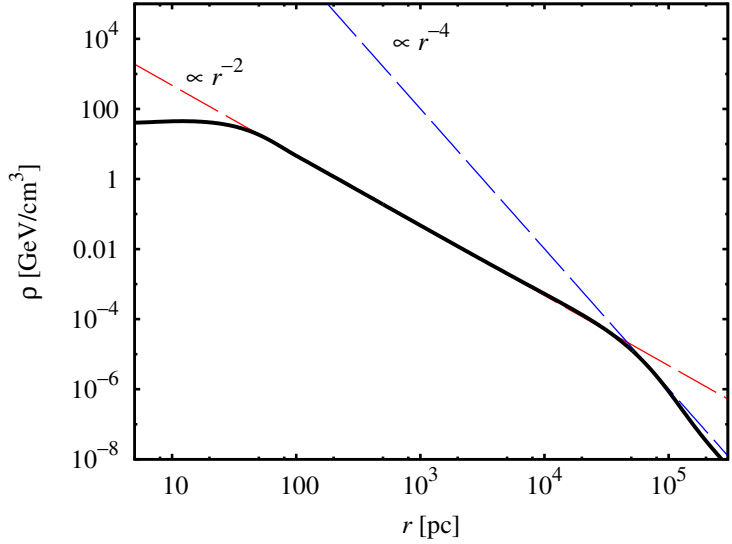}
\caption{The energy density of the $Q=1$ monopole solution with 
  $\lambda/g^2=5\times 10^6$ which corresponds to $f\sim 1.74$. Here
  we have taken $r_1=60$ pc and $r_2=30$ kpc. } 
\label{fig:energydensity}
\end{center}
\end{figure}

In Fig.~\ref{fig:energydensity} we compare the energy density of this
estimate with that of a numerical solution to the equations of motion
of the $Q=1$ monopole with $\lambda/g^2=5\times 10^6$.

\subsection{Fitting the two scalar parameters of the model}

The two parameters of the scalar sector of the model, $v$ and
$\lambda$, can be determined in terms of observable features of the
minimal dwarf at radii much smaller than $r_2$, so that screening may
be neglected and this determination will be independent of the
particular screening model.    The gauge sector is not relevant in the
core of a dark matter halo, as all components of the gauge field are
so light that their de Broglie wavelength is much bigger than the
core.

More precisely, the core density \eqref{eq:Ansatz_energydensity} is
dominated by the Higgs potential and the contribution from the
covariant derivative (see Eq.~\eqref{eq:energydensities})
\beq
\rho_{r<r_1} = \frac{699}{2240}\lambda v^4 \sim \frac{1}{3}\lambda v^4
\, ,
\eeq
where we have approximated the mean value of the core density by
setting $r=r_1/2$ in Eq.~\eqref{eq:energydensities}. The core radius
$r_1$ is given by Eq.~\eqref{raggi}. 
These two equations can easily be solved to yield the scalar sector
parameters in terms of observables 
\beq
v \simeq 0.61 \times r_1\sqrt{\rho_{r<r_1}} \hsp
\lambda \simeq \frac{23.9}{r_1^4\rho_{r<r_1}} \ .
\eeq
As a check, which is more or less guaranteed to work by dimensional
analysis, one can use the Virial theorem in the intermediate region to
verify that the stellar dispersion velocity $u$ is correctly
reproduced by these parameters 
\beq
v\sqrt{2\pi G_N}=u\sqrt{3} \, , \label{controlla}
\eeq
where $G_N$ is Newton's constant.

Segue I is the darkest known dSph \citep{segueuno} and the only galaxy
whose half-light radius is much less than $r_1$.  This means that the
corresponding velocity dispersion can be used to estimate its halo
density $\rho$ at $r<r_1$.  The dispersion has been calculated in
\citet{SegueI} to within about a factor of 2.  We will estimate
$\rho_{r<r_1}$ to be  
\beq
\rho_{r<r_1} = 30\rm{\ GeV/cm}{}^3 \, ,
\eeq
which is compatible with both the measurements of Segue I and the
dSphs in \citet{0703308}.  Furthermore we will estimate
\beq
r_1=60 {\rm \ pc} \, , 
\eeq
which is compatible with \citet{0703308} and the curves of
\citet{walker}.
Using these estimates we find the scalar parameters
\beq
v = 9 \times 10^{13}\; {\rm GeV}\hsp
\lambda=10^{-95}\, . \label{vvalore}
\eeq
It is perhaps remarkable that these galactic scale inputs, $r_1$ and
$\rho_{r<r_1}$, conspire so as to give a particle physics scale
output.  Indeed $v$ is about the GUT scale.  In the intermediate
region, if we estimate the smallest dSph stellar dispersion relation
to be 
\beq
u=6 \; {\rm km/s} \, ,
\eeq
based on, for example, \citet{walker} then it can be seen that the
check Eq.~\eqref{controlla} is satisfied to within a factor of 2. 

\subsection{Estimating $v$ using dSph masses within 300 pc} 

Now that we have fixed both parameters of the scalar sector of the
theory, we can attempt to rederive them from different data, as a
consistency check.  In this subsection we will rederive the scalar $v$
using the observation in \citet{Kap1000} that the smallest dwarf
galaxies contain $10^7$ solar masses within their innermost
300 pc. 

This derivation is not entirely independent of the
previous derivation, as the Virial theorem already relates the density
and stellar velocity dispersion.  However it has the advantage that it
does not require an estimate of $r_1$, which is difficult to determine
as the solutions of the Jeans equation for the core velocities are
degenerate if a deviation from spherical symmetry is allowed
\citep{0703308}.  We will use the fact, from the velocity curves, 
that 300 pc is much larger than $r_1$, however this will
not greatly affect our result.    Using the scaling relations derived
in Sec.~\ref{Q}, the fact that the masses vary by about a factor of
three suggests that the charge $Q$ of these galaxies varies from $Q=1$
to $Q=3$.  If we imagine that the mass $10^7\ M_\odot$ corresponds to
$Q=2$, then the corresponding mass for $Q=1$ halos will be about
$5\times 10^6\ M_\odot$.

First, note that at distances beyond $r_1$ the mass per radius is
reasonably constant.  This region yields the largest contribution to
the mass.  Thus if $M(r)$ is the mass out to a distance $r$, we will
make the approximation 
\beq
\frac{dM(r)}{dr} \sim \frac{M(r)}{r} \, ,
\eeq
in this region.  Using the figures quoted above, we then find  
\beq
\frac{dM(r)}{dr} 
\sim \frac{5\times 10^6 M_\odot}{300 \, {\rm pc}}
= 10^{29} \; {\rm GeV}^2 \, .
\label{nano}
\eeq
In this regime the gauge field is negligible and the
scalar is at its minimum.  The vast majority of the energy comes
from the winding of the scalar field, which contributes to the kinetic 
energy.  For a $Q=1$ minimal dwarf galaxy, it winds once around the
2-sphere.  In each winding the VEV moves a distance $2\pi v$, over a
physical distance $2\pi r$.  Therefore the norm of the derivative of
$\Phi$ is simply $v/r$, yielding a kinetic energy density of 
\beq
\frac{dM(r)}{dr} 
\sim 4\pi r^2 \left|\frac{\p\Phi}{\p x}\right|^2
= 4 \pi v^2 \, . \label{veq}
\eeq
Combining Eqs.~\eqref{nano} and \eqref{veq} we may now determine the
VEV $v$ 
\beq
v \sim 10^{14} \, {\rm GeV} \, ,
\eeq
again yielding a VEV which is roughly the GUT scale, in agreement with the value quoted in Eq.~\eqref{vvalore}.

\subsection{Estimating $r_2$ and $g$} \label{gvalore}

Estimating the gauge field $g$ is much more difficult, as the gauge
field only becomes relevant at distances of order $r_2$, where the
screening mechanism may also be relevant.  We will estimate it as
follows.  At large distances observations dictate that galaxies
interact largely gravitationally.  Therefore, however the screening
works, it must be subdominant to gravity at large distances.  This
implies that the energy of a galaxy must be more or less linear with
respect to its charge, so that not too much binding energy or
repulsive energy exists.  The energy density in the intermediate
region is proportional to $1/r^2$, therefore integrating over the
angular directions it is constant.  This means that the total energy
of the intermediate region, which dominates over the core energy, is
proportional to $r_2$ times the density.  We will argue below using a
simple topological argument that the density is proportional to $Q$.
The total energy is then proportional to the product of $r_2$ and $Q$
which, in order to eliminate long range gauge interactions between
galaxies, must be proportional to $Q$.  This means that $r_2$ must be
roughly $Q$ independent.  

This is one of the most robust and surprising predictions of our
model, and may well be key to its eventual falsification.  It implies
that the dark matter halos of dwarf galaxies have the same radii as
dark matter halos of LSBs, that there is a universal halo radius for
dark matter dominated galaxies.  Thus dSph dark matter halos extend an
order of magnitude beyond their half-light and tidal radii.   

So just how big is $r_2$?  Since there are so few stars at these
distances, it is difficult to estimate.  In the case of much larger
galaxies, gravitational lensing gives some estimate of the total mass
which can be used to more or less determine the extent of the dark
matter halo.  Such estimates are hard to come by for dark matter
dominated galaxies, but for the sake of concreteness one may take an
order of magnitude estimate 
\beq
r_2=30{\rm\ kpc.}
\eeq
Then Eq.~\eqref{raggi} yields
\beq
g=2\times 10^{-51}\, . \label{gvaloreeq}
\eeq

\subsection{The temperature at which the symmetry breaks}

The gauge symmetry breaks at approximately the temperature equal to
the VEV of the Higgs field, $T\sim v \sim 10^{14}\,{\rm GeV}$. This is
because when the temperature gives an effective mass to the Higgs
of this order, the symmetry is restored.

In a future publication we will attempt to study the time dependence
of the temperature of this model, assuming that it was initially in
equilibrium, due perhaps to gravitational interactions, with the
standard model fields.   For now we simply note that the fact that the
energy scale is so much higher than an eV implies that the symmetry
broke long before the recombination, and so in time to help
amplify CMB fluctuations.  The monopoles themselves could not yet form
when the gauge symmetry broke because their radius is larger than the
Hubble radius at the time.   The Hubble radius only became larger than
$r_2$ shortly before the last scattering surface, and so it is
unlikely that fully formed monopoles could have yet existed.  However
it surpassed $r_1$ much earlier, thus in these models one could have
expected the gauge symmetry breaking to have yielded some structure in
the primordial plasma.   Thus in our model it may be important that
the size of galactic cores today is smaller than the Hubble radius of
the last scattering surface and crucially also smaller than the
Hubble radius at the epoch of matter-radiation equality. 

CMB fluctuations reliably describe the primordial plasma at multipole
numbers up to about $l=$ 1,500 corresponding to fluctuations which
were about 5 kpc across at the time of recombination.
Smaller sized features affect larger multipole numbers where they are
drowned out by silk damping and the integrated Sachs-Wolfe effect.  As this size is
much larger than $r_1$, it is possible that many monopole nuclei, up
to a million, may have lied inside of each such region.   As the CMB
power spectrum is not sensitive to small enough scales ($l\gg$ 1,500)
to detect the internal structures of these monopoles, and as they
indeed have a negligible speed of sound due to their high mass, it is
likely that the monopole contribution to CMB fluctuations is
indistinguishable from that of CDM particles.   Furthermore, as they
may form when the Universe is only of order 100 years old, and as they
move very slowly, their distance from the baryon perturbations at
recombination would be more or less equal to that of CDM and thus one
could expect the same successful production of the 150 Mpc scale of large scale structure from baryon acoustic oscillations.

Note that while the various uncertainties in the parameters are only
of about 1 or 2 orders of magnitude, the model satisfies several
reasonably tight constraints.  We have remarked that it is important
that the Higgs VEV is much greater than 1 eV in order to help amplify
perturbations in the primordial plasma.  Also it is essential that the
VEV be less than the Planck scale, because otherwise gravitational
interactions would dominate over gauge interactions and so destroy the
density profiles of these monopoles, leaving instead a hairy black
hole \citep{LNW,LNW2,LNW3,BFM,BFM2}.  It is remarkable that the value of $v$
determined just by the dimensions of galaxies happens to lie in this
apparently unrelated window.

\section{Large $Q$} \label{Q}

So far we have considered only the basic monopole $Q=1$, which
corresponds in our model to the dark matter halo of the smallest
allowed dSphs.  The analysis of monopoles at higher $Q$ is complicated
by the fact that, without the fermionic sector, they are unstable.  We
will begin in Subsec.~\ref{scaling} by providing scaling arguments
which describe these unstable solutions in the absence of screening.
We will see that the cores of these solutions are stable, and so it is
not necessary that the stabilizing screening mechanism affect the
solutions at $r\ll r_2$.  However, without screening, we find that
$r_2\propto\sqrt{Q}$ whereas stability implies that $M\propto Q$ which
implies that $r_2$ is reasonably $Q$ independent.  Therefore we
conclude that the screening shifts the location of the outer radius
$r_2$ dramatically.  This in turn implies that the energy density
beyond $r_2$ is proportional not to $Q^2$ as in the unscreened case,
but to $Q$.  Thus the screening cancels the effective charge $Q$
except for a residual charge of order $\sqrt{Q}$, as might be expected
for example from a Gaussian distribution of screening antimonopoles. 

\subsection{Scaling arguments} \label{scaling}

We will assume spherical symmetry in the following argument and also that
the field strength tensor scales proportionally to $Q$.   
Furthermore, we will consider the case in which $Q$ is large and
$\lambda/g^2$ is large but finite.  It is known that no monopoles of
charge $Q>1$ are truly spherically symmetric \citep{Weinberg:1976eq},
for example monopoles of charge $2$ have only axial symmetry
\citep{Ward:1981jb}. 
However, we expect that spherical symmetry will be recovered in the
limit $Q\to\infty$.   This is reasonable in light of the restoration
of spherical symmetry in the BPS case seen in \citet{ricci,ricci2}.

Relying on the spherical symmetry we can now
assume that the scalar profile function will scale as $h\sim
r^{\daleth(Q)} \to 0$, in the region $r<r_1$ as $Q\to\infty$,
where $\daleth(Q)$ is a monotonically increasing function
\citep{Bolognesi:2005rk}. 
As the magnetic charge goes to infinity, we can neglect the
transitional regime near $r_1$ where $h$ passes from 0 to 1.  All that
matters is the size of the different regions.
We do however need to take the angular derivatives into consideration
in this argument, because the norm of the Higgs VEV will be $v$ in the
region from $r_1<r<r_2$ and its direction will wind $Q$ times 
around the vacuum manifold $S^2$. This winding will of course be
canceled by the gauge field at distances larger than $r>r_2$. 
To a good approximation these angular derivatives contribute $Q (v k
h/r)^2$ to the kinetic energy.  As will be explained in
Subsec.~\ref{sec:lambda_estimate}, this energy density  is
proportional to $Q$, instead of $(Q^2 + 1)$ as in the axial symmetric
case, as a result of the approximate spherical symmetry at large $Q$.  
Neglecting the radial derivatives we arrive at the following crude
estimate for the energy density 
\begin{align}
\mathcal{H} = 
\frac{Q^2}{2g^2r^4}\left(1-k^2\right)^2
+Q\left(\frac{v k h}{r}\right)^2
+\frac{\lambda v^4}{4}\left(1-h^2\right)^2 \, .
\label{eq:largeQ_energydensity}
\end{align}
We will estimate the functions $h,k$ to be
\beq
h(r) = 
\begin{cases}
0 \, , & \quad r \in [0, r_1] \, , \\
1 \, , & \quad r \in (r_1, \infty) \, ,
\end{cases} \qquad
k(r) = 
\begin{cases}
1 \, , & \quad r \in [0, r_2] \, , \\
0 \, , & \quad r \in (r_2, \infty) \, . 
\end{cases}
\eeq

Substituting the above estimates into the energy density
\eqref{eq:largeQ_energydensity} gives us 
\beq
M = \frac{4 \pi v}{g} 
\left[\frac{Q^2}{2 g v r_2} + Q g v \left(r_2 - r_1\right)
+ \frac{\lambda g v^3}{12} r_1^3 \right] \, .  \label{meq}
\eeq
Minimization with respect to $r_{1,2}$ then yields
\beq
r_1 = \frac{2}{v}\sqrt{\frac{Q}{\lambda}} \, , \qquad
r_2 = \frac{1}{g v}\sqrt{\frac{Q}{2}} \, , \label{r}
\eeq
which we reinsert into Eq.~\eqref{meq} to obtain
\beq
M = \frac{4 \pi v}{g} 
\left[\sqrt{2} - \frac{4g}{3\sqrt{\lambda}} \right] Q^{\frac{3}{2}} \,
. 
\eeq
Note that, as the exponent of the magnetic charge is $3/2>1$, the
monopoles will repel each other and the bound states will be unstable
in the absence of screening and gravitational attraction.    As
described in Subsec.~\ref{gvalore}, if screening leads to a
$Q$-independent value of $r_2$ then the mass will be proportional to
$Q$, and so the long distance gauge interactions of halos will be
negligible as desired. Let us now summarize our estimate of the energy
density profile 
\begin{align}
\mathcal{H}(r) = 
\begin{cases}
\frac{\lambda v^4}{4} \, , & \quad r \in [0, r_1] \, , \\
\frac{Q v^2}{r^2} \, , & \quad r \in (r_1, r_2] \, , \\
\frac{Q^2}{2g^2r^4} \, , & \quad r \in (r_2, \infty) \, . 
\end{cases}
\end{align}
Again this approximate energy density is not continuous, but it
captures the scaling in the core and intermediate regimes, although
stability implies that the scaling in the outer regime will need to be
dramatically altered by the screening. 

While an instability resulting from an energy surplus in the outer
regions of  the solution is worrying, implying the necessity of a
rather arbitrary screening mechanism, an energy surplus in the core
would be fatal.  If the core of the solution is itself unstable, then
no additional physics could stabilize these solutions without
qualitatively changing them everywhere. 

In order to make a crude estimate of the stability of the
core, we use the energy density \eqref{eq:largeQ_energydensity} to
calculate a mass up to a certain distance $r'>r_1$
\beq
M(r') = \frac{4 \pi v}{g}
\left[Q r' - \frac{4 g Q^{\frac{3}{2}}}{3\sqrt{\lambda}}\right] \, . \label{dueterm}
\eeq
The last term in this mass formula can be interpreted as a binding
energy. As this formula is only valid for $r>r_1$, the mass is
always strictly positive.  

Recall that we have assumed that the screening mechanism imposes that
$r_2$ be $Q$ independent.  Then we can determine whether the interior
is stable by simply taking the second derivative with respect to the
charge $Q$ of the mass at $r<r_2$ 
\beq
\frac{d^2 M(r')}{d Q^2} 
= - \frac{4\pi v}{g}\frac{g}{\sqrt{\lambda Q}} \, .
\eeq
The negativity of this second derivative implies that the core of the
monopole is stable and in fact even slightly bound\footnote{It would
  be interesting to determine whether this binding has observable
  consequences for galactic mergers and elliptical galaxies.}.  The
instability found above comes instead from the outer regions, where we
postulate that it will be remedied by screening.

It is clear from Eq.~(\ref{dueterm}) that our halo profiles are at best reliable up to some maximum value of $Q$. Consider a mass scaling corrected by a screening mechanism
\beq
M = \frac{4 \pi v}{g} 
\left[\sqrt{2} - \frac{4g}{3\sqrt{\lambda}}\sqrt{Q}\right] Q \,
,
\eeq
which is monotonically increasing up to the maximum charge
\beq
Q_{\rm max} \sim \frac{\lambda}{2g^2} \sim 2 \times 10^6\, .
\eeq
This charge corresponds to a mass of order $10^{15} M_\odot$, greater than that of any known galaxy and much greater than that of any dark matter dominated galaxy.

\subsection{A consistency check using larger dwarfs}
\label{sec:lambda_estimate}

So far we have essentially used three pieces of kinetic information,
the sizes $r_1$ and $r_2$ together with the core density, to determine
3 unknowns: $g$, $\lambda$ and $v$.  All of the other data that we
used could be determined from the continuity of the density function
together with the $1/r^2$ density scaling in the intermediate region.
Thus, while our model nontrivially gave the correct scalings and
satisfied some necessary inequalities, the kinetic data itself
consisted of the same number of data points and unknowns and so the
existence of a solution was reasonably trivial. 

However, now that we have determined all three parameters in the
bosonic sector of our model, which we have argued is all that is
relevant at $r<r_2$, any new kinetic data will provide a nontrivial
check of our model.  In this subsection we will consider the
intermediate region of large dwarfs, with $Q\gg 1$.  First we will
derive the scaling of the energy density in this region with respect
to $Q$.   Then we will use the data from \citet{NFW} to determine
the stellar velocity rotation curves and radii $r_1$ of such a dwarf
galaxy.  The velocity dispersion will be used to determine $Q$.  Then
we can use the fact that $r_1\propto\sqrt{Q}$ and the value of $r_1$
for a minimal dwarf to compare the value of $r_1$ predicted by our
model with the measured value.

$Q$ is easily determined from the rotational velocity by generalizing
Eq.~\eqref{veq}.  For large $Q$, 't Hooft-Polyakov monopoles are
approximately spherically symmetric.  The Higgs field winds $Q$ times
around the color $S^2= {\rm SU}(2)/{\rm U}(1)$, and so the determinant
of its derivative matrix is equal to $Q$ times that of the $Q=1$
monopole. The derivative matrix is $2\times 2$, therefore isotropy
implies that each derivative in an angular direction is enhanced by a
factor of $\sqrt{Q}$.  Thus the kinetic term and so the mass is
proportional to $Q$.  Using the Newtonian formula  
\beq
\frac{G_N M(r)}{r^2}=\frac{u^2}{r} \, , \label{vnewt}
\eeq
where $G_N$ is Newton's constant, the fact that the left hand side is
proportional to $Q$ implies that the rotational velocity $u$ is
proportional $\sqrt{Q}$.     In Subsec.~\ref{scaling} we have already
used this result to conclude that $r_1\propto\sqrt{Q}$.  Therefore we
arrive at a prediction of our model: the core radius $r_1$ is
proportional to the stellar rotational velocity in the intermediate
region. 

For concreteness, we consider the rotation curve of the dwarf galaxy
DD0168 in \citet{NFW}, although that of the other dwarf galaxies
in this reference would give a similar result.  In the intermediate
region $r_1<r<r_2$ the stars in this galaxy rotate at about 
$60 \, {\rm km}/{\rm s}$, and the inner radius is about 
\beq
r_1\sim  600 \, {\rm pc} \, . \label{ddcore}
\eeq

In the case of a $Q=1$ galaxy,  we have claimed that Eq.~\eqref{vnewt}
would give a velocity equal to the dispersion velocity of the smallest
dSph, about $6\,{\rm km}/{\rm s}$.  Since DD0168 rotates 10 times
faster, its dark matter halo consists of a monopole with charge $Q\sim
100$. As $r_1\sim\sqrt{Q}$, this implies that the radius $r_1$ of the
core of DD0168 should be about 10 times larger than that of a minimal
dwarf galaxy $Q=1$.  Above we have very roughly estimated that such
dwarfs have a core radius of 60 pc, therefore we conclude
that the core radius of DD0168 is 600 pc, in agreement with the measured value~\eqref{ddcore}.  Note that had our model
predicted that $r_1$ scales as another power of $Q$, for example were
it $Q$ independent or proportional to $Q$, then there would have been
an order of magnitude discrepancy in $r_1$.  Thus our model not only
correctly produces the density scaling as a function of radius, but
also the $Q$ dependence of the structure and density passes a
nontrivial check. 

One may be tempted to push this relation yet further, using the Milky
Way.  The core of the Milky Way is not dark matter dominated, and so a
degree of caution is needed.   The rotational velocity is about 220
km/s, suggesting $Q=$ 1,350.  Therefore one expects that
$r_1=$ 2 kpc.  This is difficult to check, the core of our
Milky way is thought to be an elliptical bar.  2 kpc
indeed may well be between lengths of the semi-minor and semi-major axes,
and so again this is consistent.  Were $r_1$  $Q$-independent, the resulting core radius of 60 pc would be
strongly excluded, as would the 80 kpc result were $r_1$
proportional to $Q$.  Thus the proportionality of $r_1$ and the
stellar rotational or dispersion velocity works quite well over this
large range of values of $Q$, whereas any other integral exponent is
very strongly excluded.

\section{Gravitating monopole} \label{gravita}

't Hooft-Polyakov monopoles have been extensively studied in the
Einstein-Yang Mills-Higgs theory
\citep{Bais:1975gu,Cho:1975uz,VanNieuwenhuizen:1975tc}.  It was
found that, when the order parameter $v$ exceeds a scale roughly equal
to the Planck energy, the monopoles become hairy black holes
\citep{LNW,LNW2,LNW3,BFM,BFM2}.  The exact threshold and the direction of the
intermonopole force depend on the ratio $\lambda/g^2$ \citep{HKK,HKK2} and
the charge $Q$ \citep{Bolognesi:2010xt}.  
We have seen that $v$ is several orders of magnitude below the Planck
scale, and thus we may expect magnetic interactions to dominate over
gravitational interactions, except for intermonopole interactions in
which the former are screened.

Thus, at least for $r<r_2$, we do not expect gravitational corrections
to our profiles to be significant.  As a check, in the $Q=1$ case, we
have used our nongravitational solutions to create a classical
Newtonian gravitational potential
\beq
V(r') = - G_N \int_{\mathbb{R}^3} 
\frac{\mathcal{H}(r)}{\left|r'-r\right|} d^3 r \, ,
\eeq
where $G_N=6.7\times 10^{-39} \, {\rm GeV}^{-2}$. Since
$\mathcal{H}(r)$ contains a factor of $v^4$, it is clearly
important that the factor $G_N v^2\ll 1$ or equivalently that 
$v\ll M_{\rm Planck}$. 
Thus the potential
will not affect our flat space solution significantly. 

As a more rigid check of the influence of the gravity on our monopole
solution, we will consider the coupling of general relativity with our
gauge theory. The Lagrangian is \citep{VanNieuwenhuizen:1975tc}
\begin{align}
\mathcal{L} &= \mathcal{L}^E + \mathcal{L}^M \, , \\
\mathcal{L}^E &= -\frac{1}{16\pi G_N}\sqrt{-g} R \, , \\
\mathcal{L}^M &= -\sqrt{-g}\Tr\left[
  \frac{1}{2g^2}F_{\mu\nu}F^{\mu\nu} +
  \D_\mu\Phi\D^\mu\Phi\right]
  -\frac{\lambda}{16\pi}\sqrt{-g}
  \left(v^2-2\Tr\left[\Phi\right]^2\right)^2 \, ,
\end{align}
where $g$ is the determinant of the metric while the Lorentz indices
are raised and lowered with the metric.
For the spherical case $Q=1$, we can assume that the metric is also
spherically symmetric and hence choose
\beq
g_{\mu\nu} = \diag\left(-e^\alpha,e^\beta, r^2, r^2\sin^2\theta\right)
\, ,  
\eeq
with $\mu,\nu=t,r,\theta,\phi$. We will use the same Ansatz for the
gauge field and change the variables of the metric to the following
gravitational variables $x\equiv(\alpha-\beta)/2$ and
$y\equiv(\alpha+\beta)/2$. 

In terms of the dimensionless radial coordinate $\eta\equiv g v r$,
the equations of motion read 
\begin{align}
y' &= \Lambda\eta U_1 \, , \label{eq:diffgrav1}\\
\left[\eta\left(e^y - e^x\right)\right]' &=
\Lambda\eta^2 e^y \left(U_1 + U_2\right) \, , \label{eq:diffgrav2}\\
\left(k' e^x\right)' &= e^y\left(\frac{1}{\eta^2}
  \left(k^2 - 1\right)k + h^2 k \right) \, , \label{eq:diffgrav3}\\
\left(\eta^2 h' e^x\right)' &= e^y\left(
 2k^2 h - \frac{\lambda\eta^2 (1-h^2)h}{g^2}\right) \, , \label{eq:diffgrav4}
\end{align}
where ${}'$ is the derivative with respect to $\eta$ and we have
defined $\Lambda\equiv 8\pi G_N v^2$ as well as the following two
functionals 
\begin{align}
U_1 &\equiv \left(\frac{k'}{\eta}\right)^2 + \frac{1}{2}(h')^2 \, , \\
U_2 &\equiv \frac{1}{2\eta^4}\left(1 - k^2\right)^2 
  +\left(\frac{k h}{\eta}\right)^2 
  +\frac{\lambda}{4g^2}\left(1-h^2\right)^2 \, .
\end{align}
The asymptotic boundary conditions are
\beq
h(\infty) = v \, , \quad
k(\infty) = 0 \, , \quad
y(\infty) = 0 \, , \quad
x(\infty) = 0 \, , \quad
\eeq
while at $r\to 0$ they are
\beq
\eta\left(e^y - e^x\right) = 0 \, , \quad
k' e^x = 0 \, , \quad
\eta^2 h' e^x = 0 \, . 
\eeq
The energy density including gravity is
\beq
\mathcal{H} = g^2 v^4 e^y \left[U_1 + U_2\right] \, .
\eeq
Glancing at the differential equations
\eqref{eq:diffgrav1}-\eqref{eq:diffgrav2} it is observed that the
factor $\Lambda \sim 3.3\times 10^{-9}$ and hence the value of $y$
will always be within the numerical error of $0$. Hence we
can already conclude that gravity does not affect the shape of our 
monopole solution. A numerical solution leads to the same conclusion.

Of course a more detailed analysis of the core at various
values of $Q$ may be interesting, in case one may find that general
relativistic corrections indeed lead to supermassive black hole (SMBH)
formation.  It may well be that in this model SMBHs are formed
primarily not by accreting stars, but rather are an integral part of
the stationary solution for the dark sector.  This would mean that
they are formed by dark forces, which as $v$ is smaller than the
Planck scale are much stronger than gravity in this regime.  This
could explain how it is that SMBHs have already grown to be as large
as they are, which is quite a challenge in CDM dark matter models.  It
may also help to explain why the sizes of galactic center black holes
obey so many universal relations to other galactic characteristics,
such as the bulge mass.

\section{Discussion and outlook} \label{conclusioni}

We have proposed a new model of dark matter halos as galactic-scale
quantized solitons.  
The idea that Dirac
quantization yields very large minimal dark matter profiles of course
is quite old, appearing in the earliest cosmic string literature
\citep{kibble} and in more modern proposals for dark matter halos 
\citep{naniloro,polonia} as classical solutions of a scalar field
\citep{florida,messicovec,messico,messicoreview}.   If the scalar is
in thermal equilibrium, as may be expected due to the similarity of
galactic rotation curves, then a fit to the parameters of these
rotation curves yields a dark matter scattering cross section which is
higher than the limits placed by the bullet cluster \citep{bullet} and
so in general these models are ruled out \citep{princeton}.  One might
already suspect that due to our extremely low couplings $\lambda$
(\ref{vvalore}) and $g$ (\ref{gvaloreeq}) our model has sufficiently
little self-interaction in order to satisfy such bounds.  Indeed, the
estimate in \citet{princeton} suggests that it is sufficient for our
fields to have masses of less than about $10^{-5}$\ eV.  Our dark
sector masses are about equal to the inverse radii $1/r_1$ and
$1/r_2$, and so are about 20 orders of magnitude within this
bound\footnote{Our masses are below the lower mass limit which
  \citet{princeton} claim is necessary to reproduce the intermediate
  range density profile.  This is consistent because our intermediate
  range density is determined not by collisions,  but by the
  nontrivial topology of the scalar condensate.}. 

Our model contains two ingredients not present in interacting scalar
models.  First of all, the $1/r^2$ density dependence of the
intermediate region results from the winding of the vacuum expectation
value $v$ around the vacuum manifold SU$(2)/$U$(1)=S^2$.  This winding
is already present in a purely scalar field configuration called the
global monopole, which was postulated as a dark matter candidate in
\citet{globalm}.  However in such cases the total mass of a monopole
is generally divergent.  We cure this divergence by introducing new
gauge fields which carry dark forces, inspired by the dark force model
of \citet{nima}.  Dark forces have been applied to the cusp problem in
\citet{niel,vogelsberger} and a model with a similar particle content
to ours has been used to generate the observed baryon asymmetry in
\citet{darkbaryon1,darkbaryon2}.  This leads one to wonder whether an
extension of our model may also contribute to baryon asymmetry. 

In this note we propose that the Dirac quantization of monopoles may
explain the apparent minimum mass of dark matter halos.
Indeed, large globular clusters and small galaxies often have
equivalent stellar content, but there is a large gap between the two
sets of solutions \citep{0703308}, and we propose that this gap is the
result of Dirac quantization of a winding number of a new $S^2$-valued
condensate about the galaxy. 

These solutions unfortunately repel, and so we are forced to add a new
species of monopole, which we hope screens this repulsion at large 
distances.  Clearly this hope, that the $\mathcal{O}(1)$ Yukawa
couplings render one monopole light and reduce the
monopole-antimonopole annihilation cross-section of monopoles of
different species while not qualitatively affecting the heavy monopole, 
is the weakest link of the proposal.  While it is difficult to verify
using semiclassical methods, it is easy to falsify, and this will be
the subject of a future publication. 

The claim leads to a lot of unexpected and pleasant features.  't
Hooft-Polyakov monopoles, unlike cold dark matter, lead to cored
density profiles of galaxies, which seem to be favored
observationally.  The parameters of this model can be determined from
observations of galaxies, and despite the galactic inputs, the outputs
are of the correct scales demanded by particle physics.  
In particular, the symmetry breaking temperature is of order of the
VEV of the dark Higgs field which is about $10^{14}\,{\rm GeV}$, 
i.e.~near the GUT scale.  As it is lower than the Planck scale,
the gauge theory and scalar interactions dominate over gravity,
allowing the monopole profile to survive gravitational corrections.
Yet it is close enough to the Planck scale that with just a bit of
screening, just outside of the Milky Way, the magnetic forces do not
cause the Milky Way to repel its satellites and neighbors, and so this
construction is consistent with the existence of the local group and
with clusters of galaxies.  It would be interesting to compare results
on monopole scattering, albeit with significant and difficult to
quantify corrections from screening, with observations from collisions
of clusters such as the Bullet cluster and MACS J0025.4-1222.  Another
pleasant feature is the $1/r^2$ fall-off of the density in an
intermediate region, reproducing the familiar result that stars at
intermediate distances tend to have constant rotational speeds, or in
the case of dSphs constant velocity dispersions.  

Despite all of these pleasant features, the monopole model of galactic
dark matter halos is easily falsifiable and probably can be falsified
with the data currently available.  As long distance interactions
continue to be gravity dominated and these monopoles are much smaller
than the basic units used in structure formation simulations, this
model does not obviously lead to any new predictions for large scale
structure.  However, it may be possible to predict the abundancy of
galaxies from a simple model of symmetry breaking along the lines of
that in \citet{kibble} and compare it with the actual abundance.
Of course, such an effort will be hampered by the fact that the
agglomeration of galaxies into larger galaxies depends heavily on the
screening mechanism.  In addition the very light fields introduced in
this model, despite their equally weak couplings, may lead to any
number of cosmological problems.

Of course a model of dark matter must do more than reproduce the halo
profiles of dark matter dominated galaxies, it must for example also
reproduce those of spiral galaxies.  While the pure dark matter halos
that we have found contain central core densities which are
essentially independent of the size of the halo, the core density is
very sensitive to the baryon density.   In fact, the inclusion of
baryons can already be seen to dramatically affect the core density in
the isothermal fits of dwarfs and LSBs in \citet{swaters}.  In spiral
galaxies, whose cores are baryon dominated, one therefore expects the
core densities to diverge dramatically from the pure dark matter value
presented here.  For example in \citet{salucci1,salucci2} it is
claimed that the central density is inversely proportional to the core
radius.  Thus the dark matter galaxies of the cores of larger galaxies
are appreciably less dense than the pure dark matter cores.  While in
principle it is possible that this reduction in density results from
the outward gravitational pull of baryons that are, for example,
ejected from supernova, it seems quite plausible that it implies that
the dark sector also interacts nongravitationally with the standard
model particles.  It remains to be seen whether such interactions may
be strong enough to explain the reduced core density and yet weak
enough to satisfy the Bullet cluster bounds of \citet{bullet}. 

One very falsifiable prediction of this proposal is the discrete
galaxy spectrum.  The main strength of the proposal is the distinction
between charges $Q=0$ and $Q=1$, corresponding to globular clusters
and minimal dwarf galaxies.  $Q=2$ is an interesting case. Monopole
solutions with spherical symmetry do not exist when $Q>1$, and when
$Q=2$ they are quite elliptical.  This leads to an energy which is
higher than it would be in the spherical case, a naive estimate using
the scaling arguments presented above yields an energy in the
intermediate regions which is about $\sqrt{5}/2$ times higher than
that of two separated $Q=1$ cores, and so may suggest that $Q=2$ halos
will be unstable with any amount of screening.  Thus the next lightest
galaxies, after $Q=2$ galaxies, may well have $Q=4$ or higher.  In
conclusion, one expects a gap in the stellar velocity dispersions of
dSphs of at least a factor of about $\sqrt{2}$.  This prediction is
already on the verge of being falsified by the Sloan Digital Sky
Survey data of \citet{Strigari2000}, and hopefully will soon lead to a
falsification of the entire model.


\subsection*{Acknowledgments}

The authors are grateful for discussions with Guido D'Amico, Roberto
Auzzi, XiaoJun Bi, Stefano Bolognesi, Javier Martinez-Magan, Antonio
Marinelli and Steven Shore. In particular, we have benefited beyond
any reasonable measure from numerous invaluable discussions with
Malcolm Fairbairn.  
JE is supported by the Chinese Academy of Sciences Fellowship for
Young International Scientists grant number 2010Y2JA01. SBG is
supported by the Golda Meir foundation fund.

\end{document}